\documentclass[aps,prd,10pt,superscriptaddress,twocolumn,nofootinbib]{revtex4-1}
\usepackage{graphicx, epsfig} 
\usepackage{amsmath,amssymb,amsfonts,dsfont,mathrsfs,amsthm,mathtools}
\usepackage{bm} 
\usepackage{color}
\usepackage{physics}
\usepackage[usenames]{xcolor}
\usepackage{hyperref}
\usepackage{siunitx}
\hypersetup{colorlinks=true,urlcolor=blue,linkcolor=magenta,citecolor=blue,filecolor=blue}
\usepackage[normalem]{ulem}
\usepackage{array}
\usepackage{hyperref}
\usepackage{booktabs}
\usepackage{cancel}
\usepackage{blindtext}
\usepackage{tikzsymbols} 
\usepackage[paperwidth=620pt,top=80pt,right=60pt,bottom=80pt,left=60pt,headheight=50pt]{geometry}
\usepackage{lipsum}

\newcommand{\diff}[1]{\text{d}#1}

\newcommand{\B}{\mathcal{B}}
\newcommand{\Lag}{\mathscr{L}}

\begin{document}

\title{Inhomogeneous metrics on complex bundles in Lovelock gravity}

\author{Crist\'obal \surname{Corral}}
\email{cristobal.corral@uai.cl}
\affiliation{Departamento de Ciencias, Facultad de Artes Liberales, Universidad Adolfo Ib\'{a}\~{n}ez, Avenida Padre Hurtado 750, 2562340, Vi\~{n}a del Mar, Chile}

\author{Borja \surname{Diez}}
\email{bdiez@estudiantesunap.cl}
\affiliation{Instituto de Ciencias Exactas y Naturales, Universidad Arturo Prat, Avenida Playa Brava 3256, 1111346, Iquique, Chile}
\affiliation{Facultad de Ciencias, Universidad Arturo Prat, Avenida Arturo Prat Chac\'on 2120, 1110939, Iquique, Chile}

\author{Daniel~\surname{Flores-Alfonso}}
\email{dafa@azc.uam.mx}
\affiliation{Departamento de Ciencias B\'asicas, Universidad Aut\'onoma Metropolitana -- Azcapotzalco, Avenida San Pablo 420, Colonia Nueva El Rosario, Azcapotzalco 02128, Ciudad de M\'exico, Mexico}

\author{Nelson \surname{Merino}}
\email{nemerino@unap.cl}
\affiliation{Instituto de Ciencias Exactas y Naturales, Universidad Arturo Prat, Avenida Playa Brava 3256, 1111346, Iquique, Chile}
\affiliation{Facultad de Ciencias, Universidad Arturo Prat, Avenida Arturo Prat Chac\'on 2120, 1110939, Iquique, Chile}

\author{Leonardo \surname{Sanhueza}}
\email{lsanhueza@udec.cl}
\affiliation{Departamento de F\'isica, Universidad de Concepci\'on, Casilla, 160-C, Concepci\'on, Chile}
\affiliation{Perimeter Institute for Theoretical Physics, 31 Caroline Street North, Waterloo, Ontario N2L 2Y5, Canada}

\begin{abstract}
We consider Lovelock gravity in arbitrary, even dimensions. We find a large class of new gravitational instantons by considering extended nontrivial circle bundles over K\"ahler manifolds. Concretely, we generalize the Page-Pope metric in the presence of higher-curvature corrections of the Lovelock class. A subset of these spaces admits analytic continuation into the Lorentzian sector, producing new stationary solutions in Lovelock gravity. The geometries are fully determined by a single algebraic equation. We also obtain necessary and sufficient conditions for Lovelock-constant K\"ahler manifolds to exist in Lovelock gravity. Finally, we find a wide class of Lovelock-Maxwell solutions beyond staticity, allowing us to obtain the electrovacuum extension of these instantons. 
\end{abstract}

\maketitle

\section{Introduction\label{sec:intro}}

There are many reasons to explore higher-curvature corrections to Einstein gravity. For instance, they are ubiquitous if general relativity is treated as an effective field theory from a Wilsonian viewpoint. They also naturally appear in the low-energy limit of string theory~\cite{Zwiebach:1985uq}, modifying the black hole spectrum of higher-dimensional general relativity~\cite{Boulware:1985wk,Wheeler:1985nh,Wheeler:1985qd}. Additionally, the renormalization of matter's stress-energy tensor in curved spacetimes requires the introduction of higher-curvature terms as well~\cite{Birrell:1982ix}. In four dimensions, including quadratic-curvature terms renders general relativity renormalizable around the Minkowski background at the one-loop level~\cite{Stelle:1976gc}, at the price of introducing ghosts~\cite{Stelle:1977ry}. In the context of the gauge/gravity duality, higher-curvature terms introduce new sources on the dual conformal field theory (CFT), allowing one to study different holographic aspects beyond Einstein-anti-de Sitter (AdS) gravity~\cite{Grumiller:2009sn,Sinha:2010ai,Alishahiha:2010bw,Kwon:2011jz,Cunliff:2013en,Ghodsi:2019xrx,Ghodsi:2020qqb,Anastasiou:2021swo,Anastasiou:2025dex}. For weakened AdS$_4$ asymptotics, it has been shown that Weyl-squared conformal gravity is finite~\cite{Grumiller:2013mxa}, while unitarity is restored by assuming Neumann boundary conditions in the Fefferman-Graham expansion~\cite{Hell:2023rbf} (see also Refs.~\cite{Maldacena:2011mk,Anastasiou:2016jix,Anastasiou:2020mik}). Another interesting setup for studying holography beyond AdS is new massive gravity in three dimensions~\cite{Bergshoeff:2009hq}. A key point here is that by introducing higher-curvature terms, gravity is now endowed with local degrees of freedom, whose dynamics admit asymptotically Lifshitz and warped AdS$_3$ black holes~\cite{Ayon-Beato:2009rgu, Clement:2009gq, Clement:2009ka, Ahmedov:2010em, Brenna:2011gp, Giribet:2012rr, Donnay:2015joa, Donnay:2015iia,Bravo-Gaete:2020ftn,Caselli:2022jlo}. 

The most general higher-curvature theory following the same principles of Einstein’s gravity in a given dimension, i.e., general covariance and leading to second-order field equations,
is Lovelock's theory~\cite{Lanczos:1938sf,Lovelock:1971yv}. It is a diffeomorphism-invariant theory constructed out of the metric and derivatives thereof, whose field equations are, at most, of second order. In four dimensions, it reduces to general relativity plus the Gauss-Bonnet term. Even though the latter does not contribute to the bulk dynamics, it can play a crucial role in renormalization of asymptotically locally AdS spacetimes~\cite{Aros:1999id,Olea:2005gb,Aros:1999kt,Miskovic:2009bm,Anastasiou:2020zwc}. In higher dimensions, however, the higher-curvature corrections of the Lovelock class become dynamical without spoiling the second-order character of the field equations. This has provided a rich arena for studying black holes~\cite{Banados:1993ur,Aiello:2004rz,Cai:2006pq,Garraffo:2008hu,Dehghani:2008qr,Mazharimousavi:2008ti,Dehghani:2010kd,Canfora:2010rh,Yue:2011et,Maeda:2011ii,Camanho:2011rj,Matulich:2011ct,Farhangkhah:2014zka,Ohashi:2015xaa,Bravo-Gaete:2021hza} and their thermodynamics~\cite{Myers:1988ze,Jacobson:1993xs,Louko:1996jd,Cvetic:2001bk,Cai:2003kt,Sahabandu:2005ma,Dehghani:2005vh,Dehghani:2006dh,Khodam-Mohammadi:2008hww,Biswas:2009pe,Dehghani:2010gn,Zou:2010yr,Miskovic:2012zz,Correa:2013bza,Xu:2014tja,Frassino:2014pha,Hendi:2015psa,Prasobh:2015rje,Cadoni:2016hhd,Hennigar:2016ekz,Singh:2020xju,Hull:2021bry,Hull:2022xew}. 

In general relativity, the existence of black holes is restricted by a number of uniqueness theorems. For example, Birkhoff's classical theorem dictates that staticity follows from assuming spherical symmetry, yielding the Schwarzschild solution uniquely. Many variations of this result exist in the literature with diverse implications. In three-dimensional general relativity, cyclic symmetry determines the rotating black hole, going beyond the static solution in this case~\cite{Ayon-Beato:2004ehj}. In Lovelock gravity, the extensions of the theorem state that the warped product between an arbitrary codimension-two Riemannian manifold and a two-dimensional Lorentzian section must be static~\cite{Dotti:2005rc,Bogdanos:2009pc,Ray:2015ava}.\footnote{Birkhoff's theorem has also been proven in higher-curvature theories beyond Lovelock~\cite{Oliva:2010eb,Oliva:2010zd}.} The higher-curvature corrections, however, impose strong constraints on the geometry of the black hole horizon. Specifically, the Lovelock field equations impose that the base manifold must be Lovelock constant, namely, each term of the Lovelock series must be proportional to the transverse metric~\cite{Ray:2015ava}. This result is a generalization of what occurs in Einstein gravity---see, for instance~\cite{Gibbons:2002bh}. Moreover, such a result allows one to write the most general static solution in terms of the nontrivial roots of a Wheeler-like polynomial~\cite{Wheeler:1985qd,Wheeler:1985nh}, whose single metric function represents a one-parameter family of solutions that is uniquely determined in terms of the Lovelock's couplings.  

Even though static black hole solutions have been widely studied in Lovelock gravity, it should be said that the highly nonlinear nature of Lovelock's equation makes it difficult to find analytic rotating solutions. Despite this, numerical solutions have been found in Refs.~\cite{Brihaye:2008kh,Brihaye:2008ns}, and slowly rotating black holes were obtained in Ref.~\cite{Kim:2007iw}. Additionally, rotating black branes can be obtained by performing a Lorentz boost from the static seed metric, rendering both spacetimes locally isometric but globally inequivalent~\cite{Dehghani:2002wn,Dehghani:2003ea,Dehghani:2006cu,Hendi:2010zza}. The obstruction for obtaining fully rotating analytic black holes in Lovelock gravity can be traced back to the no-go result of Ref.~\cite{Anabalon:2009kq}. Therein, the authors showed that the presence of the Gauss-Bonnet term prevents using the powerful Kerr-Schild ansatz, which underlies almost every rotating solution. Nevertheless, such an obstruction does not apply at the Chern-Simons point of the theory, allowing one to construct a noncircular stationary metric with two rotation parameters and devoid of a Killing horizon~\cite{Anabalon:2009kq}. The same argument was recently used to obtain a rotating and accelerating solution in five-dimensional Einstein-Gauss-Bonnet gravity at the Chern-Simons point~\cite{Anabalon:2024abz}. Indeed, this curve in parameter space is quite special, as it enlarges the internal symmetry group from Lorentz to (A)dS. Even more, it relaxes the constraint that the field equations impose on the horizon's topology, admitting wormholes~\cite{Dotti:2006cp}, topological black holes~\cite{Dotti:2007az,Dotti:2010bw}, hairy black holes with Thurston horizons~\cite{Guajardo:2024hrl}, and solitons~\cite{Tapia:2024xcp}. 

Stationary spacetimes with complex line bundle structures are starkly different. Indeed, the restriction to the Chern-Simons point is completely unnecessary for their existence. Euclidean inhomogeneous metrics possessing this structure play the key role of instantons in (super)gravity and string theory. In general relativity, the thorough analysis of this family of solutions has been performed by Page and Pope in Ref.~\cite{Page:1985bq}. These spaces include the Euclidean Taub-NUT (Taub-Newman-Unti-Tamburino) metric~\cite{Taub:1950ez}, Eguchi-Hanson metric~\cite{Eguchi:1978xp,Eguchi:1978gw}, the complex projective space, and the complex hyperbolic space, among others, which are usually regarded as gravitational instantons~\cite{Hawking:1976jb,Eguchi:1979yx,Eguchi:1980jx,Lapedes:1980qw}. In particular, the Euclidean Taub-NUT metric has been crucial for constructing the Kaluza-Klein monopole~\cite{Gross:1983hb,Sorkin:1983ns}. In asymptotically locally AdS spaces, it provides a simple setup for studying holographic fluids with nontrivial vorticity~\cite{Leigh:2011au,Caldarelli:2012cm,Leigh:2012jv,Mukhopadhyay:2013gja,Kalamakis:2020aaj}. Moreover, the presence of the NUT charge sources the magnetic part of the Weyl tensor, resembling the behavior of gravitational dyons~\cite{Araneda:2016iiy,Corral:2024lva}. Its magnetic component generates a stringlike defect known as the Misner string~\cite{Misner:1963fr}, which can contribute nontrivially to conserved charges~\cite{Mann:1999pc,Mann:1999bt,Durka:2019ajz,Hennigar:2019ive,Garfinkle:2000ms,Ciambelli:2020qny,Frodden:2021ces}. The Eguchi-Hanson metric~\cite{Eguchi:1978xp,Eguchi:1978gw}, on the other hand, is a self-dual asymptotically locally Euclidean Einstein space, relevant in string theory for resolving orbifold singularities. It is usually regarded as the gravitational analog of the Belavin-Polyakov-Schwartz-Tyupkin instanton in Yang-Mills theory~\cite{Belavin:1975fg}, whose boundary is $\mathbb{S}^3/\mathbb{Z}_2=\mathbb{RP}^3$, being a lens space. Even though some of the particular limits of the Page-Pope metric~\cite{Page:1985bq}, e.g., Taub-NUT and Eguchi-Hanson spaces, have been studied in higher-curvature gravity theories, its general form in Lovelock gravity is unknown.   

In the particular limit where the Page-Pope instantons specialize to the Euclidean Taub-NUT solution, its domain of existence in Einstein-Gauss-Bonnet gravity was studied in Ref.~\cite{Dehghani:2005zm}. Even though the field equations in the presence of cubic-curvature terms are much more involved, it is possible to obtain the general solution with and without Maxwell fields~\cite{Hendi:2008wq,Corral:2019leh}. However, the general solution to arbitrary order in the Lovelock series is lacking. On the other hand, the higher-curvature generalization of Eguchi-Hanson spaces in arbitrary dimensions is known up to quadratic order in the Lovelock series and up to third order with flat horizons; the latter was only obtained in eight dimensions~\cite{Corral:2022udb}. Nevertheless, similar to the Taub-NUT case, the Eguchi-Hanson solution to arbitrary order in the Lovelock series is unknown. To address these issues, in this paper, we propose employing the metric ansatz of the Page-Pope family~\cite{Page:1985bq}. Recall that it contains, among others, the Euclidean Taub-NUT and Eguchi-Hanson spaces as particular cases. By studying the Page-Pope metrics in Lovelock gravity, we can resolve these issues. The approach we take is to build on previous results for static solutions, which provide a solution in Wheeler polynomial form~\cite{Ray:2015ava}. 

The manuscript is organized as follows: In Sec.~\ref{sec:metric}, we provide the main geometric structures, i.e. inhomogeneous metrics as $U(1)$ fibrations over K\"ahler manifolds, and fix notation. In Sec.~\ref{sec:Einstein}, we review the Page-Pope instanton solution in even-dimensional Einstein gravity. Section~\ref{sec:Lovelock} presents the dynamics of Lovelock gravity. In Sec.~\ref{sec:Borjapolynomial}, we solve the field equations and present the solution as the nontrivial roots of a generalized Wheeler-like polynomial. In Sec.~\ref{sec:Lovelock-constant}, we study a different class of inhomogeneous metrics, that is, Lovelock-constant Einstein-K\"ahler manifolds. Section~\ref{sec:Lovelock-Maxwell} extends our results by introducing dyonic fields that solve the Lovelock-Maxwell equations. Finally, in Sec.~\ref{sec:discussion}, we discuss and summarize the results. 

\section{Inhomogeneous metrics on complex line bundles\label{sec:metric}}

We focus on inhomogeneous metrics on complex line bundles in $D=2m=2k+2$ dimensions. In Einstein gravity, these spaces were studied in Ref.~\cite{Page:1985bq} and include the higher-dimensional generalizations of the Taub-NUT and Eguchi-Hanson spaces. They are constructed as the $U(1)$ fibration of K\"ahler manifolds. Locally, they can be parametrized by the line element 
\begin{align}\label{ds}
    \diff{s^2}= \frac{\diff{r^2}}{f(r)}+f(r)h(r)(\diff{\tau}+\B)^2+N(r)\diff{\Sigma}^2\,,
\end{align}
where $\B=\B_{\bar{\mu}}\diff{x^{\bar{\mu}}}$ denotes the K\"ahler potential $1$-form that defines the symplectic structure $\Omega=\diff{\B}$ of the $2k$-dimensional K\"ahler manifold, whose metric is
\begin{align}\label{dSigma}
\dd\Sigma^2 = \bar{g}_{\bar{\mu}\bar{\nu}}(\bar{x})\dd\bar{x}^{\bar{\mu}}\dd\bar{x}^{\bar{\nu}} \,.  
\end{align}
Hereon, barred quantities are referred to as the $2k$-dimensional transverse section in Eq.~\eqref{dSigma}. 

To compute the curvature components, we use an orthonormal basis given in terms of the vielbein 1-forms $e^a=e^{a}_{\ \mu}\dd x^\mu$ such that $g_{\mu\nu}=\delta_{ab}e^{a}_{\ \mu} e^{b}_{\ \nu}$, with $\delta_{ab}$ the $2m$-dimensional Kronecker delta. Lowercase Latin characters represent $SO(2m)$ internal indices, which are decomposed as $a=\{0,1,A\}$, with capital Latin characters denoting $SO(2k)$ tangent-space indices of $\dd\Sigma^2$. Then, we can define orthonormal frames $\bar{e}^A=\bar{e}^{A}_{\ \bar{\mu}}\diff{\bar{x}^{\bar{\mu}}}$ associated with the $2k$-dimensional K\"ahler transverse section~\eqref{dSigma}, such that $\bar{g}_{\bar{\mu}\bar{\nu}}=\delta_{AB}\bar{e}^{A}_{\ \bar{\mu}}\bar{e}^{B}_{\ \bar{\nu}}$. Then, we choose 
    \begin{align}\label{frame}
    e^0&=\frac{\dd r}{\sqrt{f}}\,, & e^1&=\sqrt{fh}(\dd\tau+\B)\,, & e^{A}&=\sqrt{N}\,\bar{e}^{A}\,,
\end{align}
as the orthonormal noncoordinate basis of the metric~\eqref{ds}, where $f=f(r)$, $h=h(r)$, and $N=N(r)$. In this basis, the symplectic $2$-form can be spanned as $\Omega=\tfrac{1}{2}\Omega_{AB}\bar{e}^A\wedge\bar{e}^B$, whose inverse can be found from the relation $\Omega^{AC}\Omega_{BC}=\delta^A_B$. 

The components of the $SO(2m)$ connection can be obtained from the torsion-free Cartan's structure equation, that is, $\diff{e^a}+\omega^{a}_{\ b}\wedge e^b =0$. Then, using the frame basis in Eq.~\eqref{frame}, we obtain
\begin{subequations}\label{omegacomp}
    \begin{align}
    \omega^{01}&=-\frac{(fh)'}{2h\sqrt{f}}e^1\,, \\ \omega^{0A}&=-\frac{\sqrt{f}N'}{2N}e^{A}\,, \\
    \omega^{1A}&=\frac{\sqrt{fh}}{2N}\Omega^{A}{}_B\,e^B\,, \\ \omega^{AB}&=\bar{\omega}^{AB}-\frac{\sqrt{fh}}{2N}\Omega^{AB}e^1\,.
\end{align}
\end{subequations}
Here, prime denotes differentiation with respect to the radial coordinate $r$, and $\bar{\omega}^{AB}$ is the $SO(2k)$ connection defined via the structure equation $\diff{\bar{e}^A}+\bar{\omega}^{A}{}_B\wedge\bar{e}^B=0$.

The $SO(2m)$ curvature $2$-form can be computed from its definition, using the result in Eq.~\eqref{omegacomp}, that is,
\begin{align}\label{R2form}
    R^{ab} = \diff{\omega^{ab}} + \omega^{a}{}_c\wedge\omega^{cb} = \frac{1}{2}R^{ab}_{\ \mu\nu}\diff{x^\mu}\wedge\diff{x^\nu}\,.
\end{align}
The latter is related to the Riemannian tensor through the projection $R^{\lambda\rho}_{\ \mu\nu} = R^{ab}_{\ \mu\nu}E^{\lambda}_{\ a}E^{\rho}_{\ b}$, where $E_a = E^{\mu}_{\ a}\partial_\mu$ are the coframe components dual to the orthonormal frame basis; they satisfy $e^{a}_{\ \mu}E^{\mu}_{\ b}=\delta^a_b$ and $e^{a}_{\ \mu}E^{\nu}_{\ a}=\delta^\nu_\mu$. Using these definitions, direct computation of Eq.~\eqref{R2form} yields
\begin{widetext}
    \begin{subequations}\label{Rcomp}
\begin{align}
    R^{01}&=-\frac{1}{2\sqrt{h}}\left[\frac{(fh)'}{\sqrt{h}}\right]'e^0\wedge e^1-\frac{1}{2}\frac{N}{\sqrt{h}}\left[\frac{fh}{N}\right]'\Omega\,,\\ 
    R^{0A}&=-\frac{1}{4N'}\left[\frac{N'^2f}{N}\right]'\,e^0\wedge e^{A} -\frac{1}{4\sqrt{h}}\left[\frac{fh}{N}\right]'\Omega^{A}_{\ B}e^1\wedge e^B\,,\\
    R^{1A}&=\frac{1}{4\sqrt{h}}\left[\frac{fh}{N}\right]'\Omega^{A}_{\ B}e^0\wedge e^B+\frac{1}{2N}\sqrt{\frac{fh}{N}}\bar{D}_C\Omega^{A}_{\ B}e^C\wedge e^B +\frac{1}{4N}\left[\frac{fh}{N}-\frac{N'(fh)'}{h}\right]e^1\wedge e^A\,,\\ 
    R^{AB}&=\bar{R}^{AB}+\frac{1}{2N}\sqrt{\frac{fh}{N}}\bar{D}_C\Omega^{AB}e^1\wedge e^C-\frac{1}{2\sqrt{h}}\left[\frac{fh}{N}\right]'\Omega^{AB}e^0\wedge e^1-\frac{fN'^2}{4N^2}e^{A}\wedge e^B -\frac{fh}{4N^2}\Theta^{AB}_{CD} \,e^C\wedge e^D\,,
\end{align}
\end{subequations}
\end{widetext}
where $\bar{D}\Omega^{AB}=\dd \Omega^{AB}+\bar{\omega}^{A}_{\ C}\wedge \Omega^{CB}+\bar{\omega}^B_{\ C}\wedge \Omega^{AC}$ is the $SO(2k)$-covariant derivative of the symplectic $2$-form, the curvature $2$-form associated with the K\"ahler metric is $\bar{R}^{AB}= \diff{\bar{\omega}^{AB}} + \bar{\omega}^{A}_{\ C}\wedge\bar{\omega}^{CB}$, and we have defined
\begin{align}
\Theta^{AB}_{CD}:= \Omega^{AB}\Omega_{CD}+\Omega^{[A}_{\ [C}\Omega^{B]}_{\ D]}\,,    
\end{align}
satisfying $\Theta^{AC}_{BC} = \tfrac{3}{2}\delta^A_B$ and $\Theta^{AB}_{AB} = 3k$.

Once the field equations are solved, the absence of conical singularities at the fixed points (if any) is guaranteed if the Euclidean time is identified as $\tau\sim\tau+\beta_\tau$. This is a geometrical requirement that does not depend on the bulk dynamics. In the coordinates used in Eq.~\eqref{ds}, the Euclidean time period is given by 
\begin{align}\label{beta}
    \beta_\tau = \frac{4\pi}{f'(r)\sqrt{h(r)}}\Bigg|_{r=r_b}\,.
\end{align}
Here, $r=r_b$ is the largest positive real root of the polynomial $f(r_b)=0$. For analytically continued black holes, this period is interpreted as the inverse of the Hawking temperature. Then, since $\tau$ is periodic, the metric in Eq.~\eqref{ds} remains invariant under the action of $U(1)\times\mathrm{G}$\, where $\mathrm{G}$ is the isometry group of $\diff{\Sigma^2}$. Notice that we have not yet used the gauge freedom to fix the area coordinate. This will allow us to study different gravitational instantons on the same footing. In any case, we assume that, asymptotically, the area coordinate behaves as $N(r)\sim r^2$ when $r\to\infty$. 

\section{Page-Pope metric in Einstein gravity\label{sec:Einstein}}

The Einstein equations with cosmological constant in arbitrary dimensions are given by
\begin{align}\label{EFE}
    R_{\mu\nu} - \frac{1}{2}g_{\mu\nu}R + \Lambda g_{\mu\nu} = 0\,,
\end{align}
where $R_{\mu\nu}=R^{\lambda}_{\ \mu\lambda\nu}$ and $R=g^{\mu\nu}R_{\mu\nu}$ are the Ricci tensor and scalar, respectively. They can be obtained from a truncation of Lovelock theory to linear order in the curvature, that is, $\alpha_{p>1}=0$, with $\alpha_0=-2\Lambda$ and $\alpha_1=1$---see Sec.~\ref{sec:Lovelock} below. 

To solve the Einstein equations~\eqref{EFE} in arbitrary dimensions, first, we replace the metric ansatz~\eqref{ds} into Eq.~\eqref{EFE}. Then, by subtracting the $\tau\tau$ and $rr$ components, one obtains a first-order differential equation for the metric function $h(r)$. The latter can be solved analytically in terms of the area coordinate as 
\begin{align}\label{hsol}
    h(r) = \frac{N'(r)^2}{c\, N(r) + 1}\,,
\end{align}
where $c$ is an integration constant. Equation~\eqref{hsol} uniquely determines the metric function $h(r)$, as we still have the gauge freedom to fix $N(r)$, provided a reparametrization of the radial coordinate. Additionally, the transverse section's components of Eq.~\eqref{EFE} impose that the metric~\eqref{dSigma} must be Einstein-K\"ahler~\cite{Page:1985bq}. Then, the metric function $f(r)$ can be integrated from the remaining components of Eq.~\eqref{EFE}, whose solution is
\begin{widetext}
\begin{align}\label{Einsteinfsol}
    f(r) = \frac{\left(c\,N(r)+1 \right)^{\frac{3}{2}}}{k\,N^k(r)N'(r)^2}\left(\bar{R}\,W^{(1)}(r) - 2\Lambda W^{(0)}(r) - \mu \right)\,,
\end{align}
where $\mu$ is an integration constant, $\bar{R}$ is the scalar curvature of the transverse Einstein-K\"ahler manifold, and we have defined 
\begin{align}\label{Wp}
    W^{(p)}(r) &= \int^r\frac{N^{k-p+1}(\rho)N'(\rho)}{(c\,N(\rho)+1)^{\frac{3}{2}}}\diff{\rho}  =\,_2F_1\left(\frac{3}{2},k-p+2;k-p+3;-cN(r)\right)\frac{N(r)^{k-p+2}}{k-p+2}\,.
\end{align}
 \end{widetext}
Here, $_2F_1(a,b;c;z)$ is the Gauss hypergeometric function, which can defined in terms of the series~\cite{gil2007numerical}
\begin{equation}
    _2F_1(a,b;c;z)=\sum_{j=0}^\infty \frac{(a)_j(b)_j}{(c)_j}\frac{z^j}{j!}\,,
\end{equation}
with $(a)_j$ being the Pochhammer symbols, given by
\begin{align}
  (a)_j =  \frac{\Gamma(z+j)}{\Gamma(z)}=z(z+1)\cdots (z+j-1) \,,
\end{align}
for $j\in\mathbb{Z}_{>0}$, while $(a)_0=1$. The metric~\eqref{Einsteinfsol} is the two-parameter family of solutions found by Page and Pope in Ref.~\cite{Page:1985bq}, and it is conformally related to two different K\"ahler manifolds. Assuming a radial gauge where the area coordinate behaves $N(r)\sim r^2$ as $r\to\infty$, one can check that the remaining metric functions behave asymptotically as 
\begin{subequations}\label{fsolasymp}
\begin{align}\notag
f(r) &= -\frac{\Lambda r^2}{k(2k+1)} + \frac{3\Lambda }{ck(4k^2-1)}+\frac{\bar{R}}{2k(2k-1)} \\ \notag &\quad + \sum_{j=1}^{k-1}\frac{b_j\left[(2k+1)\bar{R} + 4(k+1)\Lambda c^{-1} \right]}{(cr^2)^j} \\  &\quad - \frac{\mu\,c^{3/2}}{4k\,r^{2k-1}} + \mathcal{O}(r^{-2k})\,, \\
    h(r) &= \frac{4}{c}\left(1 - \frac{1}{cr^2} + \frac{1}{c^2 r^4} - \frac{1}{c^3 r^6}\right) + \mathcal{O}(r^{-8}) \,.
\end{align}    
\end{subequations}
where $b_j$ are irrelevant numerical factors depending on the dimensionality.

One can check that the solution~\eqref{hsol} includes the Euclidean Taub-NUT and Eguchi-Hanson spaces as particular cases, provided a suitable choice of the integration constant $c$ and the area coordinate $N(r)$. For instance, the former is obtained by setting $c=n^{-2}$ and $N(r)=r^2-n^2$, while the latter from $c=0$ and $N(r)=r^2/4$. These limits have been studied thoroughly in Ref.~\cite{Page:1985bq} (see~\cite{Awad:2000gg} for the Taub-NUT case) alongside their regularity conditions. We will not discuss the local properties of this solution any further. Rather, we provide additional information about their global properties by interpreting the integration constants $\mu$ and $c$ in terms of conserved charges in asymptotically locally AdS (ALAdS) spacetimes.

\subsection{Self-duality and conserved charges in $D=4$}

For the sake of simplicity, let us focus on the four-dimensional case with $\mathbb{S}^2$ as the transverse Einstein-K\"ahler manifold. Then, $\diff{\Sigma^2}=\diff{\vartheta}^2 + \sin^2\vartheta\diff{\varphi}^2$ and $\mathcal{B}=\cos\vartheta\diff{\varphi}$. We consider negative cosmological constant $\Lambda=-3/\ell^2$, where $\ell$ is the AdS radius. Since we are interested in checking whether $\mu$ and $c$ are associated with conserved charges, from hereon, we assume these two quantities are nonvanishing.

In Einstein-AdS spacetimes, the Weyl tensor in arbitrary dimensions can be written as
\begin{align}
    W^{\mu\nu}_{\lambda\rho} = R^{\mu\nu}_{\lambda\rho} + \frac{1}{\ell^2}\delta^{\mu\nu}_{\lambda\rho}\,,
\end{align}
where $\delta^{\mu_1\ldots\mu_p}_{\nu_1\ldots\nu_p}=p!\delta^{[\mu_1}_{[\nu_1}\dots\delta^{\mu_p]}_{\nu_p]}$ is the generalized Kronecker $\delta$. In four dimensions, the dual-Weyl tensor is defined as
\begin{align}
    \tilde{W}_{\mu\nu\lambda\rho} = \frac{1}{2}\varepsilon_{\mu\nu\alpha\beta}W^{\alpha\beta}_{\lambda\rho}\,,
\end{align}
with $\varepsilon_{\mu\nu\alpha\beta}$ being the Levi-Civita tensor. Remarkably, there exists a curve in the parametric space of Eq.~\eqref{Einsteinfsol} where the solution becomes (anti-)self-dual, namely, $W_{\mu\nu\lambda\rho}=\pm\tilde{W}_{\mu\nu\lambda\rho}$, that is,
\begin{align}\label{muSD}
    \frac{\mu c^3}{8} = \pm\left(c-\frac{4}{\ell^2} \right) \,.
\end{align}
One can check that, if $c=4/\ell^2$, the solution represents a constant negative curvature space with $\mu=0$. This solution is trivially (anti-)self-dual, and it can be interpreted as the ground state of the theory. Nevertheless, if $c\neq4/\ell^2$ and Eq.~\eqref{muSD} holds, the solution can be interpreted as a gravitational instanton with (anti-)self-dual Weyl tensor.

In Einstein gravity, Ashtekar \textit{et.al.} provided a general prescription for computing conserved charges in ALAdS spacetimes~\cite{Ashtekar:1984zz,Ashtekar:1999jx}. Their formulation is equivalent to Noether-Wald~\cite{Wald:1993nt,Iyer:1994ys,Wald:1999wa} and quasilocal charges~\cite{Brown:1992br,Balasubramanian:1999re}, as shown in Ref.~\cite{Hollands:2005wt}.\footnote{In $D=4$, AMD charges are also equivalent to those obtained by topological renormalization~\cite{Aros:1999id,Olea:2005gb,Aros:1999kt,Miskovic:2009bm,Anastasiou:2020zwc}.} To compute the electric and magnetic Ashtekar-Magnon-Das (AMD) charges, one needs to define the electric and magnetic parts of the Weyl tensor; they are, respectively given by
\begin{align}
    \mathscr{E}_{\mu\nu} = W_{\mu\alpha \nu\beta}n^\alpha n^\beta \;\;\;\; \mbox{and} \;\;\;\; \mathscr{B}_{\mu\nu} = \tilde{W}_{\mu\alpha \nu\beta}n^\alpha n^\beta\,,
\end{align}
where $n^\mu$ is a unit-normal vector to the hypersurfaces of constant $r$. If $\xi^\mu$ is a Killing vector, then, the electric and magnetic AMD charges are defined as
\begin{subequations}\label{QAMD}
\begin{align}\label{Qelectric}
    \mathcal{Q}[\xi] &= -\frac{\ell}{8\pi G}\int_{\Sigma_\infty}\mathscr{E}^\mu_\lambda \xi^\lambda n^\nu \diff{\Sigma_{\mu\nu}} \,, \\
    \label{Qmagnetic}
    \tilde{\mathcal{Q}}[\xi] &= -\frac{\ell}{8\pi G}\int_{\Sigma_\infty}\mathscr{B}^\mu_\lambda \xi^\lambda n^\nu \diff{\Sigma_{\mu\nu}} \,,
\end{align}    
\end{subequations}
where $\diff{\Sigma_{\mu\nu}}$ is the oriented volume element of the codimension-two hypersurface $\Sigma_\infty$. Assuming that the area coordinate behaves as $N(r)\sim r^2$ as $r\to\infty$, the AMD charges for the Page-Pope instanton in ALAdS spaces are
\begin{subequations}
    \begin{align}
    \mathcal{Q}[\partial_\tau] &:= M = \frac{\mu c}{4G}\,, \\ \tilde{\mathcal{Q}}[\partial_\tau] &:= \tilde{M} = \frac{2}{c^2G}\left(c-\frac{4}{\ell^2} \right) \,,
\end{align}
\end{subequations}
whereas those associated with the Killing vector $\partial_\varphi$ vanish. Therefore, we conclude that $\mu$ and $c$ are related to the AMD charges, and they can be interpreted as the electric and magnetic mass of the Page-Pope instanton, respectively. Even more, at the (anti-)self-dual curve of Eq.~\eqref{muSD}, one can check that these conserved charges satisfy $M=\pm\tilde{M}$, rendering its electric/magnetic duality manifest.

\section{Lovelock gravity\label{sec:Lovelock}}

Lovelock gravity is the natural higher-curvature generalization of general relativity in $D>4$, as their field equations are of second order, and it propagates the same number of degrees of freedom. Although it can be formulated in arbitrary dimensions, here we focus on the even-dimensional case, recall, $D=2m=2k+2$. Its action principle is constructed solely in terms of the metric and derivatives thereof through a particular combination of the Riemann tensor, that is,
\begin{align}\label{Lovelockaction}
    I_{\rm L}[g_{\mu\nu}] = \sum_{p=0}^{k+1}\int_{\mathcal{M}}\diff{^{2m}x}\sqrt{|g|}\,\alpha_p\,\Lag^{(p)}\,,
\end{align}
where $g=\det g_{\mu\nu}$ is metric determinant, $\alpha_p$ are the Lovelock couplings, and the $p$th Lagrangian is given by
\begin{align}\label{Lagp}
    \Lag^{(p)} = \frac{1}{2^p}\delta^{\mu_1\ldots\mu_{2p}}_{\nu_1\ldots\nu_{2p}}R^{\nu_1\nu_2}_{\mu_1\mu_2}\ldots R^{\nu_{2p-1}\nu_{2p}}_{\mu_{2p-1}\mu_{2p}}\,,
\end{align}
with the normalization $\Lag^{(0)}=1$. In the even dimensions, as we are considering here, the last term of the Lovelock series is the $2m$-dimensional Euler density $\mathcal{E}_{2m}=\Lag^{(m)}$. The latter is related to the Euler characteristic of the manifold through the Chern-Gauss-Bonnet theorem. Although $\Lag^{(m)}$ does not contribute to the bulk dynamics, its presence is crucial for renormalizing the Lovelock gravity action in asymptotically locally AdS solutions~\cite{Aros:1999id,Olea:2005gb,Aros:1999kt,Miskovic:2009bm,Anastasiou:2020zwc}. However, additional counterterms are needed if the boundary is not conformally flat; they can be obtained from the conformal renormalization scheme~\cite{Anastasiou:2020mik,Anastasiou:2023oro}. 

The field equations can be obtained by performing arbitrary variations of the Lovelock action~\eqref{Lovelockaction} with respect to the metric, giving $E_{\mu\nu}=0$, where
\begin{align}\label{EOMLovelock}
    E^\mu_\nu := -\sum_{p=0}^{k} \frac{\alpha_p}{2^{p+1}}\,\delta^{\mu\mu_1\ldots\mu_{2p}}_{\nu\nu_1\ldots\nu_{2p}}R^{\nu_1\nu_2}_{\mu_1\mu_2}\ldots R^{\nu_{2p-1}\nu_{2p}}_{\mu_{2p-1}\mu_{2p}} \,. 
\end{align}
These are second-order field equations satisfying the contracted Bianchi identity, that is, $\nabla^\mu E_{\mu\nu}=0$. Additionally, its trace is related to the Lovelock Lagrangian through 
\begin{align}
    E^\mu_\mu = - \sum_{p=0}^{k} \alpha_p(m-p)\Lag^{(p)}\,.
\end{align}

In even dimensions, a curve exists in the parameter space where the theory admits degenerated maximally symmetric vacua with maximal multiplicity $k$. In that case, the Lovelock theory can be written in a Born-Infeld form~\cite{Banados:1993ur,Troncoso:1999pk,Hassaine:2016amq}. Along this curve, there is an obstruction to the linearization of the theory, where linearized methods for obtaining conserved charges give vanishing charges~\cite{Camanho:2013pda,Fan:2016zfs}. Indeed, at the nonlinear level, it was shown that conserved charges are proportional to the degeneracy condition when the maximally symmetric vacuum is unique~\cite{Arenas-Henriquez:2017xnr,Arenas-Henriquez:2019rph}, rendering the charges equal to zero at that point. We will not consider the Born-Infeld case here, as the higher-dimensional instantons we study are not continuously connected to maximally symmetric spaces. Moreover, imposing the Born-Infeld condition might induce naked singularities when the transverse section is nonconformally flat, rendering the solutions pathological. However, this is not the case if arbitrary coefficients are considered, as we do here. Hence, in what follows, we will not assume any vacuum degeneracy condition, and we will work on the general case with arbitrary Lovelock coefficients.

\section{Inhomogeneous metrics on complex line bundles in Lovelock gravity\label{sec:Borjapolynomial}}

To obtain the higher-curvature generalization of the Page-Pope metric in Lovelock gravity, we insert the ansatz in Eq.~\eqref{ds} into the field equations~\eqref{EOMLovelock}. Then, the $\tau\tau$ and $rr$ components are
\begin{subequations}
\begin{align}\label{eomtt}
   E^\tau_\tau &= -\frac{1}{2} \sum_{p=0}^k\,\alpha_p\,E^{\tau(p)}_\tau = 0\,, \\ \label{eomrr}
   E^r_r &= -\frac{1}{2} \sum_{p=0}^k\,\alpha_p\,E^{r(p)}_r = 0\,,
\end{align}    
\end{subequations}
respectively, where $E^{\tau(p)}_\tau$ and $E^{r(p)}_r$ are given by
\begin{subequations}
    \begin{align}
\notag
  E^{\tau(p)}_\tau &= 2p\,\delta^{A_1\ldots A_{2p-1}}_{B_1\ldots B_{2p-1}} R^{0B_1}_{0A_1}R^{B_2B_3}_{A_2A_3}\cdots R^{B_{2p-2}B_{2p-1}}_{A_{2p-2}A_{2p-1}} \\ & +\delta^{A_1\ldots A_{2p}}_{B_1\ldots B_{2p}} R^{B_1B_2}_{A_1A_2}\cdots R^{B_{2p-1}B_{2p}}_{A_{2p-1}A_{2p}}\,, \\
\notag
  E^{r(p)}_r &= 2p\,\delta^{A_1\ldots A_{2p-1}}_{B_1\ldots B_{2p-1}} R^{1B_1}_{1A_1}R^{B_2B_3}_{A_2A_3}\cdots R^{B_{2p-2}B_{2p-1}}_{A_{2p-2}A_{2p-1}} \\ & +\delta^{A_1\ldots A_{2p}}_{B_1\ldots B_{2p}} R^{B_1B_2}_{A_1A_2}\cdots R^{B_{2p-1}B_{2p}}_{A_{2p-1}A_{2p}}\,.
\end{align}
\end{subequations}
Here, the relevant components of the Riemannian curvature can be extracted from Eq.~\eqref{Rcomp}; they are 
\begin{subequations}
    \begin{align}
    R^{0A}_{0B} &=-\frac{1}{4N'}\left[\frac{N'^2f}{N}\right]'\delta^A_B\,, \\
    R^{1 A}_{1 B}&=\frac{1}{4N}\left[\frac{fh}{N}-\frac{N'(fh)'}{h}\right]\delta^{A}_{B}\,,\\
    R^{AB}_{CD}&=\frac{1}{2N}\bar{R}^{AB}_{CD}-\frac{fh}{4N^2}\Theta^{AB}_{CD}-\frac{fN'^2}{8N^2}\delta^{AB}_{CD}\,.
\end{align}
\end{subequations}
Subtracting Eqs.~\eqref{eomtt} and~\eqref{eomrr}, the result can be written as
\begin{align}\notag
        0 &= \sum_{p=0}^{k} \alpha_p\, p(k-p+1)\frac{fh}{N'}  \left(\frac{N'^2-h}{Nh}\right)'  \\ &\times \delta^{A_1\ldots A_{2p-2}}_{B_1\ldots B_{2p-2}}R^{B_1B_2}_{A_1A_2}\cdots R^{B_{2p-3}B_{2p-2}}_{A_{2p-3}A_{2p-2}} \,. \label{TT-RR}
\end{align}
The vanishing of the second parenthesis leads to a first-order equation for the metric function $h(r)$, whose solution is exactly that given in Eq.~\eqref{hsol}. Then, the whole system reduces to an ordinary first-order differential equation for the metric function $f(r)$. The latter admits a first integral, allowing us to express the solution of $f(r)$ in terms of the real roots of the Wheeler-like polynomial 
\begin{widetext}
\begin{align}\label{Borjapolynomial}
  \sum_{p=0}^{k} \alpha_p\left[4(k-p+1)\frac{N(r)^{k-p+2}U(r)}{(cN(r)+1)^\frac{3}{2}}B^{(p-1)}(r)+c_p\,\bar{\Lag}^{(p)}\,W^{(p)}(r)\right]=\mu\,.
   \end{align}
Here, $\mu$ is an integration constant, $c_0=1$ and $c_{p>0}=p^{-1}$, $\bar{\Lag}^{(p)}$ denotes the $p$th Lovelock density of the transverse metric~\eqref{dSigma}, $W^{(p)}(r)$ is given in Eq.~\eqref{Wp}, and we have defined 
\begin{align}
  B^{(p)}(r)&:=\sum_{l=0}^p\binom{p+1}{l}\frac{1}{2^l}U(r)^{p-l}\,\delta^{B_1\cdots B_{2p}}_{A_1\cdots A_{2p}}\,\bar{R}^{A_1A_2}_{B_1B_2}\cdots \bar{R}^{A_{2l-1}A_{2l}}_{B_{2l-1}B_{2l}}\Phi^{A_{2l+1}A_{2l+2}}_{B_{2l+1}B_{2l+2}}\cdots \Phi^{A_{2p-1}A_{2p}}_{B_{2p-1}B_{2p}}\,. \\
  U(r)&:=-\frac{f(r)N(r)'^2}{4N(r)}\,, \;\;\;\;\; \Phi^{AB}_{CD} := \frac{1}{c\,N(r) + 1}\,\Theta^{AB}_{CD} + \frac{1}{2}\delta^{AB}_{CD}\,.
\end{align}
\end{widetext}
where $B^{(-1)}=0$ and $B^{(0)}=1$ are understood. Additionally, the field equations along the components of the base manifold are satisfied as long as $\diff{\Sigma^2}$ is a Lovelock-constant Einstein-K\"ahler manifold. 

We have checked the validity of Eq.~\eqref{Borjapolynomial} explicitly up to fourth order in the Lovelock series, and it reproduces all the well-known solutions and beyond. For instance, the Page-Pope solution~\cite{Page:1985bq} of Einstein gravity is recovered by setting $\alpha_0=-2\Lambda$, $\alpha_1=1$, and $\alpha_{p>1}=0$. Moreover, the Wheeler-like polynomial~\eqref{Borjapolynomial} also includes the static solution of Ref.~\cite{Ray:2015ava} if $c=0$ and $\mathcal{B}=0$, generalizing the latter to stationary cases. On the other hand, the Taub-NUT solutions found in Refs.~\cite{Dehghani:2005zm,Hendi:2008wq,Corral:2019leh} are generalized to arbitrary order in the Lovelock series by choosing $N(r)=r^2-n^2$ and $c=n^{-2}$. In higher-curvature theories beyond Lovelock, the Taub-NUT/Bolt metric has been studied in Ref.~\cite{Bueno:2018uoy}. Additionally, the higher-curvature generalization of the Eguchi-Hanson metric found in Ref.~\cite{Corral:2022udb} is also extended by choosing $N(r)=r^2/4$ and $c=0$. Therefore, the generalized Wheeler's polynomial in Eq.~\eqref{Borjapolynomial} contains the whole family of higher-curvature corrections of the Lovelock class to the well-known gravitational instantons in Einstein gravity. This is the main result of our paper. 

Since we are not considering curves in parameter space where maximally symmetric vacua are degenerated, the asymptotic behavior of this solution resembles that of Eq.~\eqref{fsolasymp} but with an effective cosmological constant that depends on the different $\alpha_p$ couplings. We do not display its explicit value here as it is cumbersome and not illuminating. However, the falloff of the mass parameter $\mu$ is unaffected by the higher-curvature corrections, contrary to what happens with the leading-order terms.

\section{Lovelock-constant Einstein-K\"ahler manifolds\label{sec:Lovelock-constant}}

Let us discuss a different class of solutions to the field equations~\eqref{EOMLovelock}. These correspond to Lovelock-constant Einstein-K\"ahler manifolds, meaning that each term of Eq.~\eqref{EOMLovelock} must be proportional to the metric (see Ref.~\cite{Ray:2015ava}). Notice that, for $D=2k$ with $k\geq2$, the only flat Einstein-K\"ahler space is $\mathbb{T}^{2k}$, while all the others have nonconstant Riemannian curvature. This is the case, for instance, of $(\mathbb{S}^2)^k$, $(\mathbb{H}^2)^k$, $\mathbb{CP}^k$, and $\mathbb{CH}^k$. Since $\mathbb{T}^{2k}$ is flat, Eq.~\eqref{EOMLovelock} is trivially satisfied as long as $\alpha_0=0$. However, the Lovelock field equations need to be analyzed carefully for nonconstant curvature Einstein-K\"ahler spaces. 

The line element and K\"ahler potential $1$-form of $\mathbb{CP}^k$ are given by the iterative formula
\begin{widetext}
\begin{subequations}\label{CPK}
\begin{align}
 \mathcal{B}_{(k)} &= (k+1)\sin^2\psi_k\left(\diff{\phi_k} + \frac{1}{k}\mathcal{B}_{(k-1)}\right),\label{BkCPK}\\
 \diff{\Sigma_{(k)}^2} &= 2(k+1)\bigg[\diff{\psi_k^2} + \sin^2\psi_k\cos^2\psi_k\left(\diff{\phi_k} + \frac{1}{k}\mathcal{B}_{(k-1)} \right)^2 
 + \frac{1}{2k}\sin^2\psi_k\diff{\Sigma_{(k-1)}^2} \bigg]\,, 
 \end{align}
\end{subequations}
\end{widetext}
where $0\leq\psi_k\leq\pi/2$ and $0\leq\phi_k\leq 2\pi$. The case for $k=1$ reduces locally to that of $\mathbb{S}^2$. However, no hyperspheres can be obtained from $\mathbb{CP}^k$ if $k\geq2$. On the other hand, the complex hyperbolic space, $\mathbb{CH}^k$, can be obtained by taking the trigonometric functions of $\psi_k$ in Eq.~\eqref{CPK} and replacing them with their hyperbolic counterparts, while keeping invariant the trigonometric functions of $\psi_{j<k}$. In that case, $-\infty<\psi_k<\infty$ but all else is as in Eq.~\eqref{CPK}. Then, replacing the metrics of $\mathbb{CP}^k$ and $\mathbb{CH}^k$ into the Lovelock equations, we find that they are satisfied if the condition 
\begin{align}\label{polconstraintCPCH}
    \sum_{p=0}^{k}\frac{\alpha_p}{(k-p)!(k-p+2)!}\left(\frac{2\gamma}{k+2}\right)^{p-1}=0\,,
\end{align}
is met, where $\gamma=\pm1$ is related to the Gaussian curvature of $\mathbb{CP}^k$ and $\mathbb{CH}^k$, respectively. This imposes a relation between the $\alpha_p$ and $\gamma$ that provides the necessary and sufficient condition for $\mathbb{CP}^k$ and $\mathbb{CH}^k$ to solve the Lovelock equations.  

In the case of $(\mathbb{S}^2)^k$ and $(\mathbb{H}^2)^k$, on the other hand, a similar computation yields 
\begin{align}
    \sum_{p=0}^{k}\frac{\alpha_p}{(k-p)!}(2\gamma)^{p-1}=0\,,
\end{align}
where $\gamma=\pm1$ parametrize $(\mathbb{S}^2)^k$ and $(\mathbb{H}^2)^k$, respectively. Similar to the previous cases, this condition allows one to solve one Lovelock coupling in terms of the others such that these spaces solve the field equations.  

\section{Charged solutions in Lovelock-Maxwell theory\label{sec:Lovelock-Maxwell}}

In addition to the higher-curvature corrections to gravitational instantons in vacuum, some of them have been studied in the presence of Maxwell fields as well. For instance, in Ref.~\cite{Mann:2005mb}, the Taub-NUT-Reissner-Nordstr\"om solution was found in even-dimensional Einstein gravity. This solution reduces to the higher-dimensional Reissner-Nordstr\"om black hole in the limit of vanishing NUT charge. Additionally, the charged generalization of the Taub-NUT metric in Einstein-Gauss-Bonnet gravity was obtained in Ref.~\cite{Dehghani:2006aa}. Here, we present their extension to arbitrary order in the Lovelock series. 

We focus on electrovacuum solutions to the Lovelock-Maxwell theory in $D=2m=2k+2$ dimensions. Its dynamics is described by the action principle
\begin{align}
    I[g_{\mu\nu},A_\mu] &= I_{\rm L}[g_{\mu\nu}] - \frac{1}{4}\int_{\mathcal{M}}\diff{^{2m}x}\sqrt{|g|}\,F_{\mu\nu}F^{\mu\nu}\,,
\end{align}
where $I_{\rm L}[g_{\mu\nu}]$ is defined in Eq.~\eqref{Lovelockaction} and the $U(1)$ field strength is $F_{\mu\nu}=\partial_\mu A_\nu - \partial_\nu A_\mu$. The field equations for the metric and gauge potential are given by
\begin{align}\label{EOMLovelockMaxwell}
    E_{\mu\nu} =  T_{\mu\nu}  \;\;\;\;\; \mbox{and} \;\;\;\;\;
    \nabla_\mu F^{\mu\nu} = 0\,,
\end{align}
respectively, where $E_{\mu\nu}$ is defined in Eq.~\eqref{EOMLovelock} and the Maxwell stress-energy tensor is
\begin{align}\label{Tmunu}
    T_{\mu\nu} = F_{\mu\lambda}F_{\nu}^{\ \lambda} - \frac{1}{4}g_{\mu\nu}F_{\lambda\rho}F^{\lambda\rho}\,.
\end{align}

To solve the field equations~\eqref{EOMLovelockMaxwell}, we assume the line element~\eqref{ds}, together with a Maxwell field aligned along the Hopf fibration of the Einstein-K\"ahler manifold, i.e.,
\begin{align}\label{MaxwellAnsatz}
A=A_\mu\diff{x^\mu} = a(r)\left(\diff{\tau}+\B\right)\,,    
\end{align}
where $\mathcal{B}$ is the K\"ahler potential $1$-form related to the symplectic structure via $\Omega=\diff{\mathcal{B}}$ (see Sec.~\ref{sec:metric}). Then, the $U(1)$ field strength is given by
\begin{align}\label{Fansatz}
    F = \diff{A} = \frac{a'(r)}{\sqrt{h(r)}}\,e^0\wedge e^1 + a(r)\,\Omega\,.
\end{align}
Inserting the latter into the Maxwell equations, we find that the system reduces to a second-order differential equation
\begin{align}\label{MaxwellEDO1}
    \frac{1}{\sqrt{h}}\left[\frac{a'}{\sqrt{h}}\right]'  +\frac{k}{N}\left(\frac{a'N'}{h}-\frac{a}{N}\right)=0\,.
\end{align}
For the metric ansatz in Eq.~\eqref{ds}, the field strength in Eq.~\eqref{Fansatz} produces a Maxwell stress-energy tensor whose nontrivial components are
\begin{subequations}
\begin{align}
    T_r^r&=T_\tau^\tau=\frac{1}{2}\left(\frac{a'^2}{h}-\frac{ka^2}{N^2}\right)\,,\\
    T^{i}_j&=-\left[\frac{(k-2)}{2}\frac{a^2}{N^2}+\frac{a'^2}{2h}\right]\delta^{i}_j\,,\\
    T^\tau_i&=\left(\frac{a'^2}{h}-\frac{a^2}{N^2}\right)\mathcal{B}_i\,.
\end{align}    
\end{subequations}
Since $T^\tau_\tau = T^r_r$, subtracting the $\tau\tau$ and $rr$ components in the Lovelock-Maxwell equations yields exactly Eq.~\eqref{TT-RR}, whose solution is given in Eq.~\eqref{hsol}. Inserting the latter into Eq.~\eqref{MaxwellEDO1}, we obtain
\begin{align}\notag
    &\frac{\sqrt{cN+1}}{N'}\left[\frac{a'\sqrt{cN+1}}{N'}\right]'\\ & \quad +\frac{k}{N}\left(\frac{a'(cN+1)}{N'}-\frac{a}{N}\right)=0\,. \label{MaxEOMfinal}
\end{align}

Equation~\eqref{MaxEOMfinal} can be integrated analytically in terms of the functions $W^{(p)}(r)$ in Eq.~\eqref{Wp}, giving
\begin{equation}\label{asol}
     a(r)=\frac{\sqrt{cN(r)+1}}{N(r)^k}\left(q_1-q_2\,W^{(1)}(r)\right)\,,
\end{equation}
where $q_1$ and $q_2$ are integration constants. To associate these integration constants with conserved charges via a Gauss law, we first need to study the asymptotic behavior of the field strength $2$-form and its dual. The Maxwell profile in Eq.~\eqref{asol} behaves
\begin{align}
 a(r) &=  -\frac{2q_2}{c(2k-1)} + \frac{q_1\sqrt{c}}{r^{2k-1}} + \mathcal{O}(r^{-2})\,,
\end{align}
as $r\to\infty$. The leading-order term of $q_2$, however, will induce a divergence on the Gauss law associated with the Maxwell field if $k>1$. This can be seen from the asymptotic behavior of the dual field strength along the transverse components of the constant $\tau-r$ section, namely, 
\begin{align}
    \star F &= -q_2\Bigg(\frac{4k\,r^{2k-3}}{(2k-1)(2k-3)} + \mathcal{O}(r^{2k-5})\Bigg)\bar{\varepsilon}_{(2k)}\,,
\end{align}
where $\bar{\varepsilon}_{(2k)}$ is codimension-two volume $2k$-form and the contribution of $q_1$ does not induce any divergence whatsoever. Therefore, if $k>1$, the regularity of the Maxwell field implies that $q_2=0$; its contribution is finite if and only if $k=1$. Otherwise, it is divergent in general higher dimensions. Hence, once the regularity condition $q_2=0$ is imposed, $q_1$ is a conserved charge defined by the Gauss law
\begin{align}
    Q_e := \frac{1}{\text{vol}(\Sigma)} \int_\Sigma\star F = \frac{(2k-1)}{2}c\,q_1\,,
\end{align}
where $\Sigma$ is a codimension-two boundary, whose volume is denoted by $\text{vol}(\Sigma)$. 

If $k=1$, regularity conditions do not force $q_2$ to vanish. In such a case, one can associate $q_2$ with a magnetic charge via the codimension-two integral of the first Chern class, that is,
\begin{align}
    Q_m := \frac{1}{\text{vol}(\Sigma)}\int_\Sigma F = -\frac{2q_2}{c}\,.
\end{align}
Therefore, in four dimensions, the Maxwell solution can be interpreted as an Abelian dyon. 

In four dimensions, there exists a curve in parameter space where the Abelian dyon becomes (anti-)self-dual, namely, $F_{\mu\nu}=\pm \tilde{F}_{\mu\nu}$ with $\tilde{F}_{\mu\nu}=\tfrac{1}{2}\varepsilon_{\mu\nu\lambda\rho}F^{\lambda\rho}$. The condition such that the last relation is met is 
\begin{align}
    q_2 = \mp \frac{q_1\,c^2}{4}\,.
\end{align}
Along this curve, $Q_e = \pm Q_m$ is satisfied, and the electric/magnetic duality of the Abelian (anti-)self-dual dyon becomes manifest.

Returning to the $k>1$ case with $q_2=0$ and replacing the solution~\eqref{asol} into the Lovelock-Maxwell equations, we find that the system can be analytically integrated by the nontrivial roots of the Wheeler-like polynomial
\begin{widetext}
\begin{align}\label{BorjapolynomialMaxwell}
  \sum_{p=0}^{k} \alpha_p\left[4(k-p+1)\frac{N(r)^{k-p+2}U(r)}{(cN(r)+1)^\frac{3}{2}}B^{(p-1)}(r)+c_p\,\bar{\Lag}^{(p)}\,W^{(p)}(r)\right] + P(r) =\mu\,,
   \end{align}    
where all the definitions of Sec.~\ref{sec:Borjapolynomial} have been used, $P(r)$ encodes the backreaction of the Abelian gauge field in Eq.~\eqref{asol} to the Lovelock-Maxwell equations, and it is given by
\begin{equation}
    P(r):=-\frac{1}{2}\int^r\left\{\frac{a(\rho)'^2N(\rho)^{k+1}}{N(\rho)'\sqrt{cN(\rho)+1}}-\frac{k\,a(\rho)^2N(\rho)'N(\rho)^{k-1}}{(cN(\rho)+1)^{3/2}}\right\}\dd \rho\,.
\end{equation}
\end{widetext}
To the best of our knowledge, this is the first four-parameter family of inhomogeneous solutions of the Lovelock-Maxwell system representing a stationary and charged extension of that in Ref.~\cite{Ray:2015ava} to arbitrary order in the Lovelock series. It includes, as a particular limit, different well-known charged gravitational instantons in Einstein and Lovelock gravity, extending the Page-Pope metric in the presence of higher-curvature corrections.

\section{Discussion\label{sec:discussion}}

In this paper, we have constructed a new class of gravitational instantons in Lovelock gravity. They exist in arbitrary but even dimensions since the geometries are complex line bundles over K\"ahler manifolds. The metric ansatz in Eq.~\eqref{ds} is the one employed by Page and Pope in Ref.~\cite{Page:1985bq} to construct inhomogeneous Einstein metrics. The order of the Lovelock series is determined by the dimension, and we consider arbitrary coupling constants throughout. 

The solutions we have presented here specialize to those found by Page and Pope when the higher-order Lovelock coupling constants vanish. Our large class of solutions contains the Eguchi-Hanson instanton~\cite{Eguchi:1978xp}; as well as previously found higher-curvature generalizations of it~\cite{Corral:2022udb}. We mention these, in particular, as they are known to be Euclidean metrics, which cannot be rotated into Lorentzian geometries. A complementary subset of solutions are those of Taub-NUT type, which can be analytically continued into spacetimes with Lorentzian signature. The latter also possesses an interesting static limit corresponding to black holes with K\"ahler horizon geometries.

In the static limit, our class of gravitational instantons reduces to solutions found by Ray in Ref.~\cite{Ray:2015ava}, modulo a change in signature. Ray's class is not constrained by spacetime dimensionality, nor does it require that the base/transverse manifold have a K\"ahler structure. Hence, our solutions represent a nonstatic generalization of only a subclass of Ray's. To find our solutions, we have employed a strategy similar to Ray, which was established by Wheeler in Refs.~\cite{Wheeler:1985nh,Wheeler:1985qd}. The approach can be summarized succinctly as integrating the Lovelock equations such that the solution is given by a single algebraic equation whose degree coincides with the order of the Lovelock series. In previous work, it was shown that it was possible to apply this procedure to Taub-NUT spacetimes, which lack static symmetry~\cite{Corral:2019leh}. In the present work, that approach is fully extended to arbitrary dimensions after applying it to metrics with complex line bundle structures. In particular, Eq. \eqref{Borjapolynomial} shows that Wheeler's original strategy, which applies generically to static Lovelock metrics, also applies to a large class of geometries beyond that symmetry. In Ref.~\cite{Corral:2019leh}, it was exhibited that the polynomial forms of static spacetimes are not appropriate for describing stationary metrics. For instance, deformations of the polynomial terms were considered generically insufficient for the task. This issue is explicitly resolved by Eq.~\eqref{Borjapolynomial}.

Toward the end of the paper, we couple Maxwell fields to Lovelock gravity. By considering electromagnetic fields aligned with the complex null directions of the metrics, we obtain new Lovelock electrovacua. This class of charged solutions is also shown to possess Wheeler polynomial form, cf. Eq.~\eqref{BorjapolynomialMaxwell}. Moreover, the solution is simply given by the vacuum polynomial with an additional term that fully represents the backreaction of the Abelian field on the geometry. Hence, these results lead us to at least two conclusions. First, the preliminary results of~\cite{Corral:2019leh,Corral:2022udb} are seen to apply for all Taub-NUT and Eguchi-Hanson spaces for the entire family of Lovelock gravities. This is to say, all previously known solutions of those types are special cases of Eq.~\eqref{BorjapolynomialMaxwell}, and all new solutions of those types are provided by it as well. Second, all gravitational instantons and Lorentzian spacetimes, which share their geometrical structure, show similar dynamics under the Lovelock equations.

Interesting questions remain open. For instance, the conserved charges and thermodynamics of these spaces are certainly worth exploring. Nevertheless, as they have generically nonconformally flat boundaries, dealing with their divergencies in ALAdS spaces is nontrivial. 
Additionally, since the asymptotic boundary of these configurations possesses a rich structure, their holographic properties would provide an interesting setup for studying strongly coupled systems on nontrivial background geometries. Indeed, it would be interesting to check whether the partition function of the dual CFT develops similar universal behavior to those found on squashed spheres~\cite{Bueno:2018yzo}. We postpone these questions for future works.

\begin{acknowledgments}
We thank Andr\'es Anabal\'on, Giorgos Anastasiou, Ignacio J.~Araya, Luis Avil\'es, Eloy Ay\'on-Beato, Fabrizio Canfora, Oscar Fuentealba, Gast\'on Giribet, Mokhtar Hassa{\"{i}}ne, Rodrigo Olea, and Julio Oliva for insightful comments and remarks. We also thank Sim\'on del Pino for participating in the early stages of this work. L. S.  is grateful to Luca Ciambelli for his
hospitality at Perimeter Institute during the final stages of this work. The authors acknowledge the partial support of Agencia Nacional de Investigaci\'on y Desarrollo (ANID), Chile, via Fondecyt Regular Grants No. 1240043, No. 1240048, No. 1251523, No. 1252053, No. 1230112, 1231133. B.D. is supported by Becas de Mag\'ister UNAP. D. F. A. acknowledges financial support from SECIHTI through a postdoctoral research grant. L. S.  is supported by Beca Doctorado Nacional ANID Grant No. 21221813 and Beneficios Complementarios de Pasantía ANID.  This research was supported in part by the Perimeter Institute for Theoretical Physics. Research at Perimeter Institute is supported by the Government of Canada through the Department of Innovation, Science, and Economic Development, and by the Province of Ontario through the Ministry of Colleges and Universities.

\end{acknowledgments}

\bibliography{References}

\end{document}